# Tomographic active optical trapping of arbitrarily shaped objects by exploiting 3-D refractive index maps

Kyoohyun Kim[1,2] and YongKeun Park[1,2,3,*]

[1]Department of Physics, Korea Advanced Institute of Science and Technology (KAIST), Daejeon 34141, Republic of Korea

[2]KI for Health Science and Technology (KIHST), KAIST, Daejeon 34141, Republic of Korea 3TomoCube Inc., Daejeon 34051, Republic of Korea

*Correspondence:: yk.park@kaist.ac.kr

**Optical trapping can be used to manipulate the three-dimensional (3-D) motion of spherical particles based on the simple prediction of optical forces and the responding motion of samples. However, controlling the 3-D behaviour of non-spherical particles with arbitrary orientations is extremely challenging, due to experimental difficulties and the extensive computations. Here, we achieved the real-time optical control of arbitrarily shaped particles by combining the wavefront shaping of a trapping beam and measurements of the 3-D refractive index (RI) distribution of samples. Engineering the 3-D light field distribution of a trapping beam based on the measured 3-D RI map of samples generates a light mould, which can be used to manipulate colloidal and biological samples which have arbitrary orientations and/or shapes. The present method provides stable control of the orientation and assembly of arbitrarily shaped particles without knowing *a priori* information about the sample geometry. The proposed method can be directly applied in biophotonics and soft matter physics.**

Optical tweezers have been an invaluable tool for trapping and manipulating micrometre-sized spherical particles. In optical tweezers, a tightly focused laser beam generates a gradient force that attracts colloidal particles and biological cells near the optical focus[1] (Fig. 1a). In the past two decades, the development of wavefront shaping techniques has facilitated the invention of holographic optical tweezers, which can simultaneously generate multiple optical foci in three-dimensions by displaying engineered holograms on a variety of diffraction optical elements[2, 3] (Fig. 1b).

The optical forces exerted on a spherical particle can be analytically calculated using Mie theory. To predict the optical forces on particles with low symmetry, however, requires extensive numerical calculations, such as the T-matrix method[4] and discrete dipole approximation[5]. Previous works have shown that non-spherical particles can be aligned along a limited equilibrium orientation when trapped with a Gaussian beam[6, 7], and have exhibited unstable motion, depending on the sample geometry and optical properties[8, 9]. Since optical trapping is an example of light-matter interaction, methods of controlling the stable orientation of arbitrarily shaped particles can be explored either by modifying sample shapes or by engineering the wavefront of light[10]. Recent advances in two-photon polymerization now enable the fabrication of arbitrarily shaped samples with trapping handles for stable orientation control[11, 12], and iterative optimization of phase-only holograms using the T-matrix calculation have provided enhanced trap stiffness for spherical particles[13]. However, controlling the stable orientation of arbitrarily shaped particles using wavefront shaping based on sample geometry has yet to be explored.

Here, we present a novel technique, called a tomographic mould for optical trapping (TOMOTRAP). TOMOTRAP provides the stable control of the orientation and shape of arbitrarily shaped samples (Fig. 1c). TOMOTRAP measures the 3-D refractive index (RI) maps of samples in real-time and generates 3-D light field distributions, whose shapes resembles the desired sample shape and orientation. According to the electromagnetic variational principle[14], the electromagnetic energy functional for arbitrarily shaped dielectric materials with permittivity distribution, $\varepsilon(\vec{r})$, in an electromagnetic field, $E_f(\vec{H})$, which is described as

$$E_f(\vec{H}) = \frac{\langle \vec{H}|\hat{H}|\vec{H}\rangle}{\langle \vec{H}|\vec{H}\rangle} = \frac{\int_V d\vec{r} |\nabla \times \vec{E}(\vec{r})|^2}{\int_V d\vec{r} \varepsilon(\vec{r}) |\vec{E}(\vec{r})|^2} \quad , \tag{1}$$

is minimized when the overlap volume between the materials and the light intensity is maximized (See Supplementary Information). For that reason, dielectric materials change their orientation and become aligned with the high-intensity gradient of optical fields. This hinders the optical control of non-spherical particles aligned in arbitrary orientations when using conventional Gaussian optical traps. Alternatively, the 3-D light intensity generated by TOMOTRAP, whose shape is identical to the arbitrarily shaped sample, will automatically maximize the overlap volume between the light and the arbitrarily orientated sample, and acts like a light mould. This leads to the stable control of the orientation of arbitrarily shaped samples having arbitrary orientations.

## Results

### Experimental setup

The TOMOTRAP concept is schematically described in Fig. 2a, and the optical setup for TOMOTRAP is shown in Fig. 2d. Initially, the 3-D RI distribution of samples is measured. Next, the wavefront of the trapping beam is calculated from the 3-D RI distribution. This calculated wavefront will give arise to a 3-D beam intensity shape that is identical to the 3-D volumes of samples, and have the same desired orientation and/or morphology as those obtained by the measured tomogram. Then, the calculated wavefront is displayed onto a sample, and this maximizes the overlapping volume between the samples and the trapping beam intensity Finally, the samples are aligned with the updated orientation and morphology in three dimensions as intended.

In order to measure the 3-D RI distribution of samples, we employed real-time optical diffraction tomography (ODT)[15] (Fig. 2b). ODT reconstructs the RI tomograms of samples from multiple spatially modulated holograms, which are recorded by a Mach-Zehnder interferometer (See Methods). Then, the desired 3-D shapes of the samples to be translated, rotated and folded were calculated by applying translation, rotation and folding transformations, respectively, to the reconstructed tomogram of the samples in their initial state.

The trapping beam was generated by implementing holographic optical tweezers, and the wavefront of the trapping beam, whose 3-D beam intensity resembles the desired 3-D shapes of the samples, was calculated by employing the 3-D Gerchberg-Saxton algorithm[16, 17]. The 3-D Gerchberg-Saxton algorithm uses iterative Fourier and inverse Fourier transforms to find a 2-D phase-only hologram which can yield the desired 3-D beam intensity located at a Fourier plane of the hologram (see Methods). To generate the desired 3-D shape of the beam intensity on the sample plane, the calculated phase-only hologram is displayed on a spatial light modulator (SLM) illuminated by a high-power laser beam. As predicted above, the samples were aligned with the updated 3-D beam intensity, which was calculated from the tomogram measurements in order to maximize the overlap volume between the sample and beam intensity. The 3-D behaviour of the samples during the alignment was measured by time-lapse ODT.

### Orientation control of arbitrarily shaped colloidal particles

In order to verify the proposed idea and investigate the 3-D behaviour of arbitrarily shaped particles in the desired 3-D beam shape, we first trapped and controlled the arbitrary orientation of a PMMA ellipsoidal dimer (Fig. 3a and Supplementary Movie 1). The colloidal PMMA ellipsoids were fabricated by heat stretching[18] 3 μm diameter colloidal PMMA spheres embedded in PVA films (See Methods).

Initially, the tomogram of a PMMA dimer in 45% w/w sucrose solution was measured, and the TOMOTRAP controlled the arbitrary orientation of the PMMA dimer by rotating the sample with respect to the $x$-, $y$- and $z$-axis. As shown in Fig. 3a, the proposed method was able to change the orientation of the PMMA dimer, even when the dimer was aligned along the optical axis.

In order to quantitatively analyse the feasibility of the proposed orientation control method, we calculated the 3-D cross-correlation values between the calculated tomogram of the desired orientation and the measured tomogram. The calculated 3-D cross-correlation values were maintained at $0.92 \pm 0.026$ during the orientation changes (Supplementary Fig. 1), which clearly shows the high feasibility of the present method for controlling the orientation of arbitrarily shaped particles.

In addition to controlling the orientation of individual arbitrarily shaped colloidal particles, the proposed method enables the simultaneous translational and rotational control of multiple particles with arbitrary shapes. This feature was used to assemble multiple PMMA particles (Fig. 3b and Supplementary Movie 2). We first trapped two PMMA dimers and a separate PMMA ellipsoid, and the translational and rotational motion of each particle was controlled independently, to assemble all of the particles into a PMMA particle complex. The assembled complex was then stably translated and rotated together.

**Orientation and shape control of biological samples**

TOMOTRAP is also capable of controlling the orientation and shape of biological samples which have more complicated geometry. As shown in Fig. 4a and Supplementary Movie 3, the present method trapped individual red blood cells (RBCs) on a cover glass from the reconstructed tomogram of the RBCs, whose initial orientation was *en face* to the optic axis. The RBCs were then sequentially rotated with respect to the $y$-axis and $z$-axis while maintaining the initial discoid shape of the RBC. After being aligned along the $z$-axis, showing edge-on orientation, the RBC was folded into an L-shape. The desired 3-D beam intensity was calculated by applying a folding transformation using the measured tomogram. The folding transformation was designed as a rotation transformation of one half of the sample, whose rotation axis was set to be a lower bisector of the sample. The measured tomogram (the last two columns in Fig. 4a) clearly shows the L-shaped RBC. The folded RBC was then rotated with respect to the $z$-axis while maintaining the L-shaped folding.

In addition to controlling the orientation and shape of individual RBCs, the present technique can also be utilized to assemble multiple biological samples (Fig. 4b and Supplementary Movie 4). Initially, two RBCs were sedimented on a cover glass with face-on orientation. The present method independently rotated each RBC with respect to the $x$- and $y$-axis of the centre of mass of each RBC, and sequentially translated the two RBCs, assembling them as a T-shaped complex of RBCs consisting of two RBCs in an edge-on orientation. The assembled RBCs were then sequentially rotated with respect to the $z$- and $x$-axis, and the final orientation of the RBCs was with one face-on, and one edge-on.

**Discussion**

In summary, we proposed and experimentally demonstrated TOMOTRAP for stably controlling the orientation and shape of arbitrarily shaped particles. Exploiting the electromagnetic variational principle, we theoretically predicted that dielectric samples would be aligned to the 3-D beam intensity of a desired shape and orientation, which acts as a tomographic mould for optical trapping. Employing an optical setup that combined ODT and holographic optical tweezers, we experimentally demonstrated that the proposed idea can control the orientation and/or shape of arbitrarily shaped particles, including PMMA ellipsoidal dimers and RBCs.

The present method provides stable control of the orientation and assembly of arbitrarily shaped particles without knowing *a priori* information about the sample geometry. This work can be applied promptly to various fields such as the 3-D assembly of arbitrarily shaped microscopic particles, including colloidal particles[19, 20], bacteria[21] and stem cells[22]. It is also noteworthy that the present method can be used to induce a desired shape in samples by mechanical deformation, which permits the 3-D optical sculpting of various materials[23]. We also anticipate that TOMOTRAP could benefit studies in biomechanics, and can be used to investigate the active microrheology of the fluctuating membranes of biological samples[24, 25] with global optical deformation, as well as the 3-D optical guiding of cellular migration[26].

## Methods

**Sample preparation.** Poly(methyl methacrylate) (PMMA) ellipsoids were fabricated by one-dimensional (1-D) heat stretching of PMMA spheres embedded in polyvinyl alcohol (PVA) films[8, 9, 18]. PMMA spheres with diameters of 3 μm (86935-5ML-F, Sigma-Aldrich Co. MO, USA) were embedded in PVA films (341584, $M_w$ = 89,000 – 98,000, Sigma-Aldrich Co.) and then mechanically stretched in a glycerol bath at a temperature of 130°C, which is higher than the glass transition temperature of PMMA ($T_g \approx$ 105 °C). After stretching, the PMMA ellipsoids were recovered by dissolving the PVA films in 20% isopropanol/water solution and by washing with the same solution several times. The PMMA ellipsoids were then immersed in 45% w/w sucrose and 0.1% w/w TWEEN-20 (P9416-50ML, Sigma-Aldrich Co.) solution. The PMMA ellipsoids in a suspension of 50 μL were loaded between two coverslips (24 × 50 mm, C024501, Matsunami Glass Ind., Ltd., Japan) spaced by two strips of double-sided Scotch tape.

All the red blood cells (RBCs) measured in our experiments were collected from healthy donors. 5 μL drops of the blood were collected from healthy volunteers by a fingertip needle prick and diluted in 1 mL of Dulbecco's Phosphate-Buffered Saline (DPBS, Welgene Inc., Korea) and 4% w/w bovine serum albumin (BSA, 30063-572, Thermo Fisher Scientific Inc. MA, USA) solution. In order to prevent adhesion of the RBCs to the coverslips, the coverslips were coated with 4% w/w BSA solution and incubated for 30 minutes. After incubation, the BSA solution was gently washed with distilled water, and a 50 μL RBC suspension was loaded between two coverslips spaced by two strips of double-sided Scotch tape.

**Optical diffraction tomography.** The three-dimensional (3-D) refractive index (RI) distribution of the colloidal and biological samples was reconstructed by optical diffraction tomography (ODT)[27, 28]. A Mach-Zehnder interferometer was used to measure the optical fields diffracted by the samples. A diode-pumped solid-state (DPSS) laser ($\lambda_l$ = 532 nm, 100 mW, Cobolt Samba, Cobolt AB, Sweden) beam was split into two arms by a beam splitter. One arm was used as a reference beam, and the other arm illuminated the samples on an inverted microscope (IX 71, Olympus Inc., Japan) through a tube lens ($f$ = 200 mm) and a water-immersion condenser lens with a high numerical aperture (NA = 1.2, UPLSAPO, 60×, Olympus Inc.). For tomographic measurements, the incident angle of the illumination beam was tilted by a dual-axis scanning galvanomirror (GVS012/M, Thorlabs Inc., NJ, USA). The galvanomirror circularly scanned 10 illumination beams with various azimuthal angles at a scanning rate of 10 msec/cycle. The diffracted beam from the samples was collected by a high NA objective lens (NA = 1.4, UPLSAPO, 100×, oil immersion, Olympus Inc.), and the beam was further magnified 4 times by an additional 4-*f* configuration. The diffracted beam interfered with the reference beam at the image plane of the samples, which generates spatially modulated holograms. Multiple holograms from various illumination angles were recorded by a high-speed CMOS camera (1024 PCI, Photron Inc., CA, USA) at a frame rate of 1,000 Hz.

Complex optical fields of the samples, consisting of amplitude and phase delay, were extracted from the recorded holograms by a field retrieval algorithm based on the Fourier transform[29]. The 2-D Fourier spectra of the retrieved complex optical fields were mapped onto the surface of a hemisphere, called an Ewald sphere, in the 3-D Fourier space based on the Fourier diffraction theorem[27]. The 3-D RI distribution of the samples was reconstructed by taking the inverse Fourier transform of the 3-D Fourier space. All processes including hologram acquisition, field retrieval, and tomogram reconstruction were performed in a custom-made MATLAB GUI interface, and the use of a graphics processing unit (GPU, GTX 970, nVidia Co., CA, USA) enabled real-time tomogram reconstruction. Reconstructing a tomogram of $128^3$ voxels ($21.8 \times 21.8 \times 21.8$ μm) for all processes took approximately 2 sec. Recently, optical diffraction tomography has been commercialized[30].

**Holographic optical tweezers.** The optic setup for holographic optical tweezers shares the same high-NA objective lens in the inverted microscope of the ODT. A high-power DPSS laser ($\lambda_T$ = 1,064 nm, 10 W, MATRIX 1064-10-CW, Coherent Inc., CA, USA) beam illuminated a spatial light modulator (SLM, X10468-07, Hamamatsu Photonics K.K., Japan). The SLM displayed phase-only holograms which modulate the wavefront of the trapping beam. By adding a phase grating pattern to the phase-only hologram on the SLM, the unmodulated (zeroth-order) beam was separated from the modulated (first-order) beam, which was blocked by a spatial filter. The first-order beam was further demagnified by an additional 4-$f$ configuration in order to overfill the back aperture of the objective lens. The detailed optical setup for combining the ODT and holographic optical tweezers can be found elsewhere[15].

The phase-only holograms for trapping arbitrarily shaped particles from the measured 3-D RI distributions were generated by implementing the 3-D Gerchberg-Saxton algorithm[16, 17]. The 3-D Gerchberg-Saxton algorithm relates 3-D beam intensity as an objective and 3-D $k$-space as a physical constraint for beam propagation by 3-D Fourier transform pairs. From the measured 3-D RI distribution of the samples, the desired 3-D beam shape was generated by applying rotational, translational and/or folding transformation to the measured tomograms. The 3-D Fourier spectra of the desired 3-D beam shape were obtained by a 3-D Fourier transform. Then, the 3-D Fourier spectra outside of the surface of the Ewald sphere of the trapping beam became zero, and the amplitude of the 3-D Fourier spectra on the surface of the Ewald sphere of the trapping beam was replaced by a constant, which conserves the total energy of the trapping beam. The modified 3-D Fourier spectra were back-propagated by taking an inverse 3-D Fourier transform to generate the updated 3-D beam shape. The amplitude part of the updated 3-D beam shape was replaced by the initial desired 3-D beam shape, and propagated to the 3-D Fourier spectra again. After repeating the iterative process 30 times, the 2-D projection of the angular part of the 3-D Fourier spectra on the Ewald sphere yields a phase-only hologram to be displayed on the SLM, which can generate the desired 3-D beam shape. By employing the GPU, the total computation time for generating a phase-only hologram from a measured 3-D RI distribution took less than 1 sec.

**Acknowledgements**
This work was supported by KAIST, and the National Research Foundation of Korea (2015R1A3A2066550, 2014K1A3A1A09063027, 2012-M3C1A1-048860, 2014M3C1A3052537) and Innopolis foundation (A2015DD126).

**Author Contributions**
K. K. designed and performed experiments. Y. P conceived the idea and supervised the project. All the authors wrote the manuscript.

**Competing Financial Interests**

The authors declare no competing financial interests

**Figures with Captions**

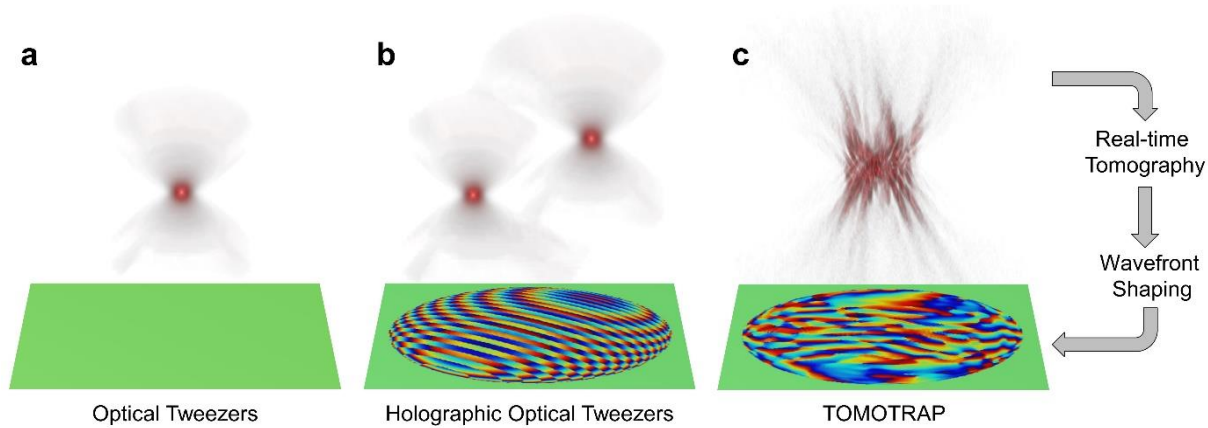

**Figure 1 | Schematic diagrams of various optical tweezers.** Schematic diagrams of **a,** single-beam optical tweezers, **b,** holographic optical tweezers, and **c,** TOMOTRAP employing real-time 3-D RI tomography and wavefront shaping. The top row depicts the 3-D beam intensity generated by each of the optical tweezers, and the bottom row shows the phase component of the complex optical field of the trapping beam.

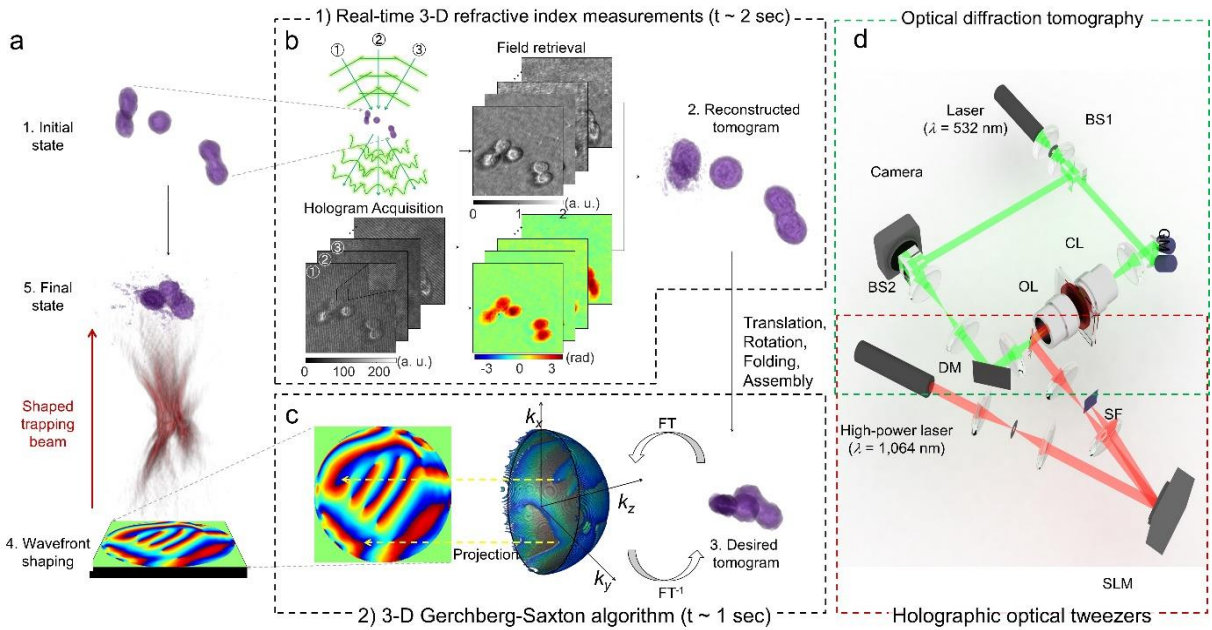

**Figure 2 | Principle of controlling the stable orientation and assembly of arbitrarily shaped particles using TOMOTRAP. a,** Schematic diagram of TOMOTRAP. **b,** Real-time optical diffraction tomography reconstructing the 3-D refractive index (RI) distribution of the samples from measured multiple holograms. **c,** The 3-D Gerchberg-Saxton algorithm, calculating a 2-D phase-only hologram to generate the desired 3-D beam intensity by iterative Fourier and inverse Fourier transforms. **d,** The optical setup for TOMOTRAP, consisting of optical diffraction tomography and holographic optical tweezers. A diode-pumped solid state (DPSS) laser ($\lambda_I$ = 532 nm) beam was split into two arms by a beam splitter (BS1). One arm was used as a reference beam, and the other arm illuminated the samples on an inverted microscope through a tube lens and a water-immersion condenser lens (CL) with high numerical aperture (NA = 1.2, 60×). For tomographic measurements, the incident angle of the illumination beam was tilted by a dual-axis scanning galvanomirror (GM). The diffracted beam from the samples was collected by a high NA objective lens (OL), and the beam was further magnified 4 times by an additional 4-*f* configuration. The diffracted beam interfered with the reference beam at the image plane of the samples, which generated spatially modulated holograms. Multiple holograms from various illumination angles were recorded by a high-speed CMOS camera. The optic setup for the holographic optical tweezers shares the same high-NA objective lens in the inverted microscope of the ODT. A high-power DPSS laser ($\lambda_T$ = 1,064 nm) beam illuminated a spatial light modulator (SLM). The SLM displayed phase-only holograms which modulated the wavefront of the trapping beam. By adding a phase grating pattern to the phase-only hologram on the SLM, the unmodulated (zeroth-order) beam was separated from the modulated (first-order) beam, which was blocked by a spatial filter (SF). The first-order beam was further demagnified by an additional 4-*f* configuration in order to overfill the back aperture of the OL.

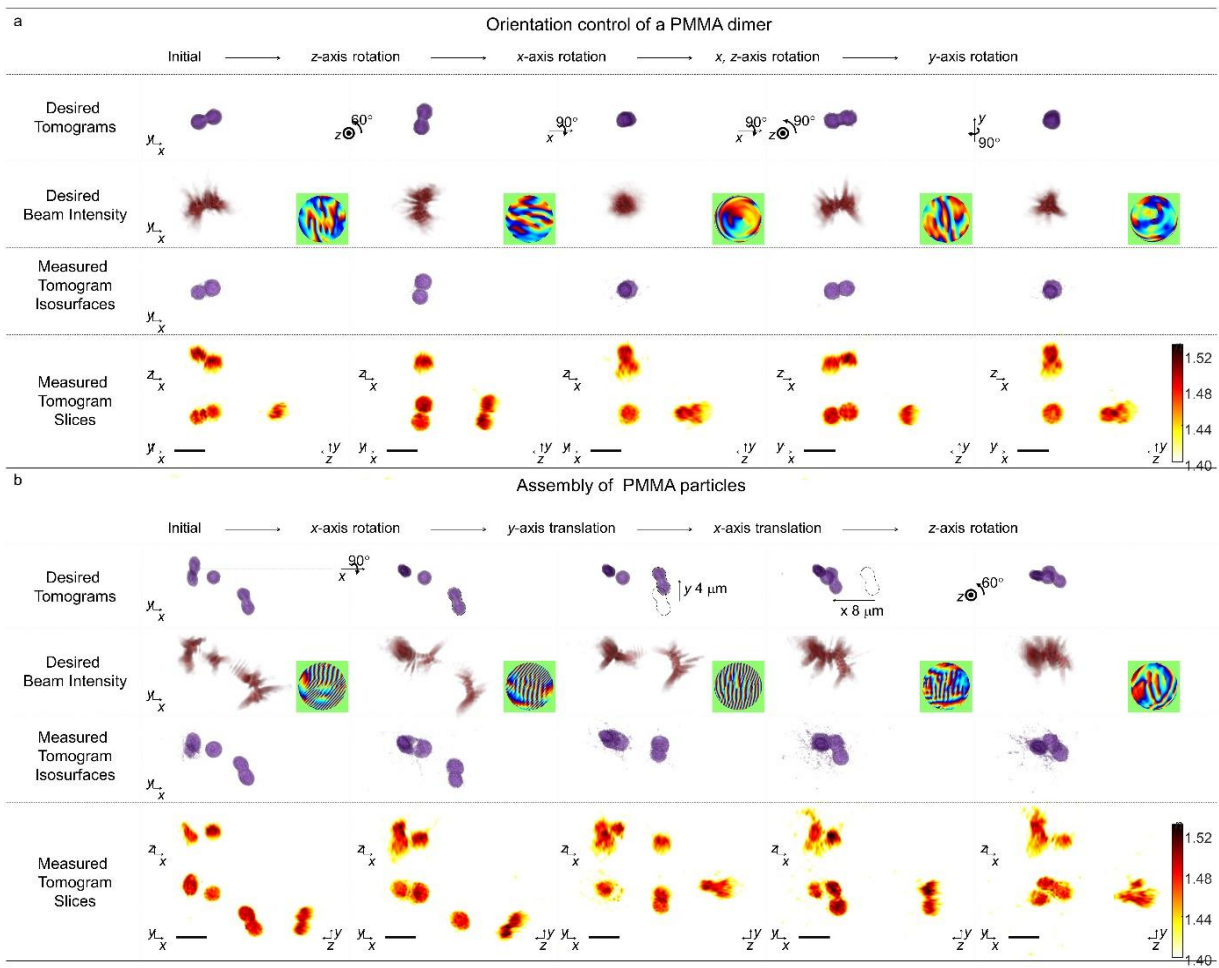

**Figure 3 | Controlling the orientation and assembly of colloidal PMMA particles. a,** Time-lapse images of the controlled orientation of a PMMA dimer. See also Supplementary Movie 1. **b,** Time-lapse images of the assembly of three PMMA particles. See also Supplementary Movie 2. First row: desired tomograms calculated by applying rotational, translational, and/or folding transformations to the reconstructed tomogram in the initial state. Second row desired 3-D beam intensity generated by numerical propagation of the phase-only hologram in the insets of each column. The phase-only holograms were calculated by applying the 3-D Gerchberg-Saxton algorithm to the desired tomograms in the first row of each column. Third row: 3-D rendered isosurfaces of the tomograms of the PMMA particles trapped by the desired 3-D beam intensity at each column of the second row. Fourth row: the cross-sectional slice images of the measured tomograms in the *x-y*, *x-z*, and *y-z* plane. Scale bar indicates 5 μm.

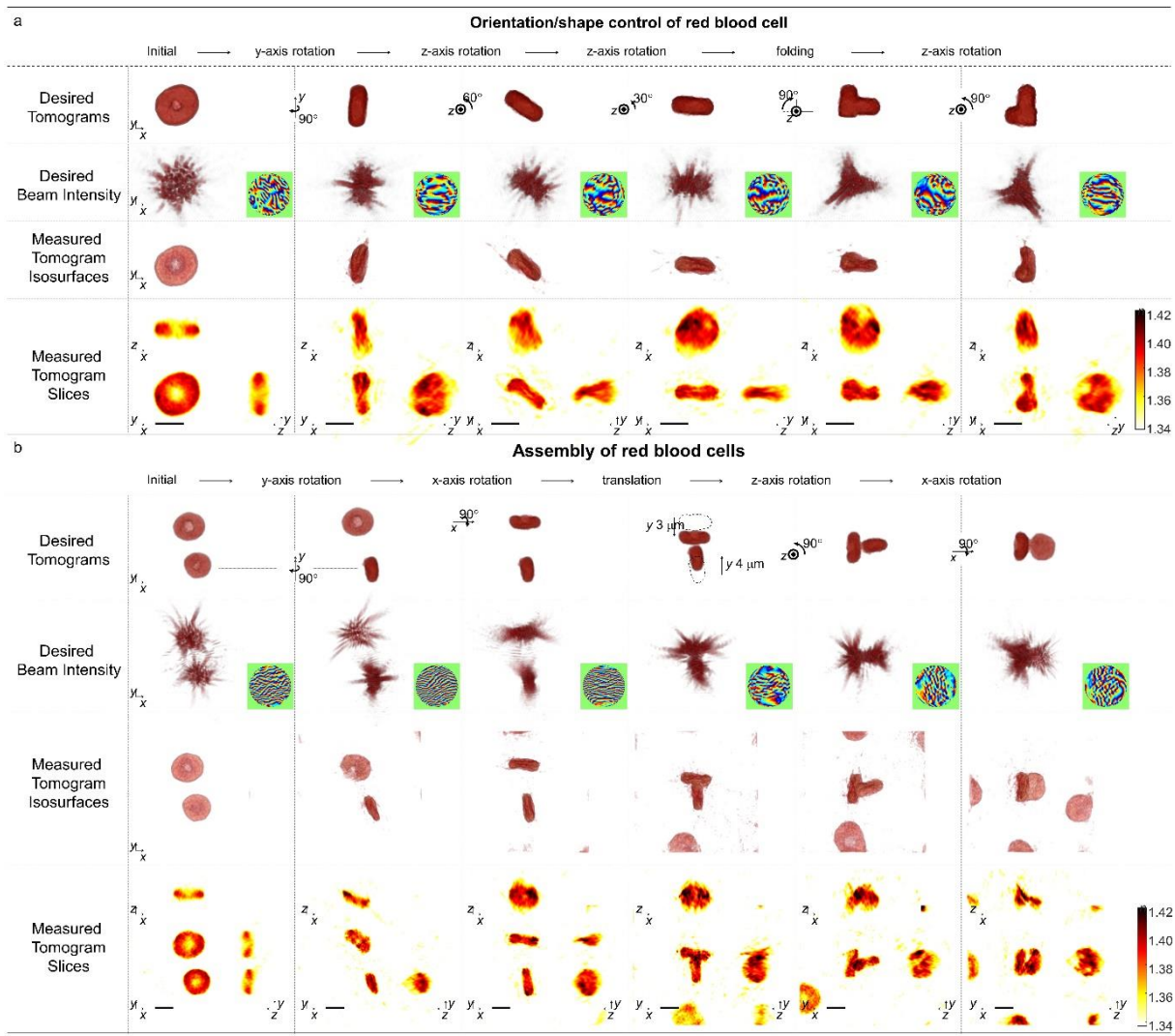

**Figure 4 | Controlling the orientation, shape and assembly of red blood cells. a,** Time-lapse images of orientation control of individual red blood cells (RBCs). See also Supplementary Movie 3. **b,** Time-lapse images of the assembly of two RBCs. See also Supplementary Movie 4. First row: desired tomograms calculated by applying rotational, translational, and/or folding transformations to the reconstructed tomogram of the initial state. Second row: desired 3-D beam intensity generated by numerical propagation of the phase-only hologram in the insets of each column. The phase-only holograms were calculated by applying the 3-D Gerchberg-Saxton algorithm to the desired tomograms in the first row of each column. Third row: 3-D rendered isosurfaces of the tomograms of RBCs trapped by the desired 3-D beam intensity at each column of the second row. Fourth row: images of the cross-sectional slice of the measured tomograms in the *x-y*, *x-z*, and *y-z* planes. Scale bar indicates 5 μm.

# Supplementary information:

## S1 : Electromagnetic variational principle

In the absence of free charges and currents, Maxwell's equations for dielectric and nonmagnetic particles are:

$$\nabla \cdot \varepsilon \vec{E} = 0, \qquad \nabla \cdot \vec{H} = 0,$$
$$\nabla \times \vec{E} = -\frac{\partial \vec{H}}{\partial t}, \quad \nabla \times \vec{H} = \varepsilon \frac{\partial \vec{E}}{\partial t}, \qquad (1)$$

where $\vec{E}(\vec{r},t)$ and $\vec{H}(\vec{r},t)$ are the electric and magnetic field, respectively, and $\varepsilon(\vec{r})$ is the permittivity distribution of the non-dispersive particles.

When we only consider harmonic time-dependent field with a harmonic frequency $\omega$,

$$\vec{E}(\vec{r},t) = \vec{E}(\vec{r})\exp(-i\omega t),$$
$$\vec{H}(\vec{r},t) = \vec{H}(\vec{r})\exp(-i\omega t) \qquad (2)$$

Maxwell's equations become

$$\nabla \cdot \varepsilon \vec{E}(\vec{r}) = 0, \qquad \nabla \cdot \vec{H}(\vec{r}) = 0,$$
$$\vec{H}(\vec{r}) = -\frac{i}{\omega}\nabla \times \vec{E}(\vec{r}), \quad \vec{E}(\vec{r}) = \frac{i}{\omega\varepsilon}\nabla \times \vec{H}(\vec{r}). \qquad (3)$$

The magnetic field, $\vec{H}(\vec{r})$, can be rewritten by substituting the electric field to the last equation as:

$$\omega^2 \vec{H} = \nabla \times \left(\varepsilon^{-1}(\vec{r})\nabla \times \vec{H}\right) \qquad (4)$$

Eq. (4) can be interpreted as an eigenvalue problem,

$$\hat{H}|\vec{H}_\omega\rangle = \omega^2 |\vec{H}_\omega\rangle,$$
$$\qquad (5)$$

where $\vec{H}_\omega$ is an eigenfunction of a Hermitian Maxwell Hamiltonian $\hat{H}$ as

$$\hat{H} = \nabla \times \left(\varepsilon^{-1}(\vec{r})\nabla \times\right). \qquad (6)$$

The expectation value of the Hamiltonian $\hat{H}$ is

$$\langle \vec{H}_\omega | \hat{H} | \vec{H}_\omega \rangle = \int_V d\vec{r} \vec{H}^*(\vec{r}) \cdot \nabla \times \left( \varepsilon^{-1}(\vec{r}) \nabla \times \vec{H}(\vec{r}) \right) = \int_V d\vec{r} \frac{1}{\varepsilon(\vec{r})} \left| \nabla \times \vec{H}(\vec{r}) \right|^2$$

$$= \int_V d\vec{r} \omega^2 \varepsilon(\vec{r}) \left| \vec{E}(\vec{r}) \right|^2 = \omega^2 \langle \vec{H}_\omega | \vec{H}_\omega \rangle \qquad (7)$$

$$= \int_V d\vec{r} \left[ -\frac{i}{\omega} \nabla \times \vec{E}(\vec{r}) \right]^* \cdot \nabla \times \left[ \frac{1}{\varepsilon(\vec{r})} (-i\omega \varepsilon(\vec{r})) \vec{E}(\vec{r}) \right] = \int_V d\vec{r} \left| \nabla \times \vec{E}(\vec{r}) \right|^2$$

The electromagnetic variational principle states that the electromagnetic energy functional

$$E_f(\psi) = \frac{\langle \psi | \hat{H} | \psi \rangle}{\langle \psi | \psi \rangle}, \qquad (8)$$

is minimized at the ground state $|\psi_g\rangle$ with the ground energy $E_g$, and the value of the functional gives the energy of the ground eigenstate. By substituting Eq. (7) to Eq. (8), the electromagnetic energy functional becomes

$$E_f(\vec{H}_\omega) = \frac{\langle \vec{H}_\omega | \hat{H} | \vec{H}_\omega \rangle}{\langle \vec{H}_\omega | \vec{H}_\omega \rangle} = \frac{\int_V d\vec{r} \left| \nabla \times \vec{E}(\vec{r}) \right|^2}{\int_V d\vec{r} \varepsilon(\vec{r}) \left| \vec{E}(\vec{r}) \right|^2}, \qquad (9)$$

which is minimized when the electric field is concentrated in volumes of permittivity distribution in order to maximize the denominator of the quotient. For that reason, dielectric materials in optical fields tend to align with the high gradient of beam intensity in order to maximize the overlap volume between the materials and beam intensity.

## S2 : Feasibility of TOMOTRAP for the orientation control of a PMMA dimer

The feasibility of TOMOTRAP for the orientation control of a PMMA dimer was quantitatively analysed by calculating the 3-D cross-correlation values between the calculated tomogram ($n_{mould}$) of the desired orientation and the measured tomogram ($n_{measured}$). Time-lapse optical diffraction tomography was used to measure 20 tomograms of the PMMA dimer during the orientation change with the tomogram acquisition rate of 100 Hz. The 3-D cross-correlation values were calculated as the maximum value of the 3-D cross-correlation of two tomograms, $\langle n_{mould}, n_{measured} \rangle$, which were calculated by applying 3-D Fourier transforms as

$$\langle n_{mould}, n_{measured} \rangle = \mathcal{F}^{-1} \left[ \left( \mathcal{F}\{n_{mould}\} \right)^* \cdot \mathcal{F}\{n_{measured}\} \right]. \qquad (9)$$

As shown in Fig. S1, the 3-D cross-correlation values were calculated as 0.92 ± 0.026 during the orientation changes and translation of the PMMA dimer. It clearly shows that the present method can control the orientation of arbitrarily shaped particles with high feasibility.

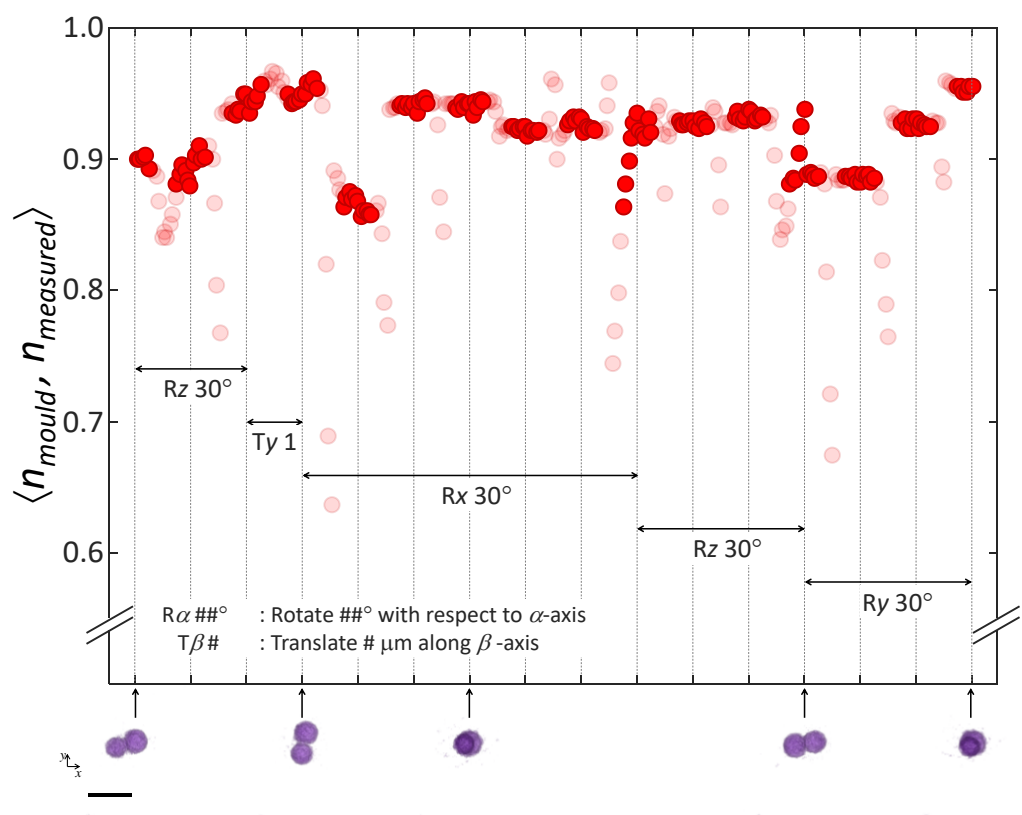

Figure S1. The 3-D cross-correlation values between the desired tomograms and the measured tomograms during the orientation control of a PMMA dimer presented in Fig. 3a. The rotation and translation transformation indicated as R$\alpha$ ##° and T$\beta$ #, respectively, were applied at every dashed line. For instance, rotation transformation for rotating the PMMA dimer 30° with respect to the $x$-axis (R$x$ 30°) was applied 6 times. Solid red circles are the 3-D cross-correlation values when the PMMA dimer is trapped steadily in the 3-D beam intensity of desired tomograms, while shaded circles indicate intermediate states during the orientation change in response to applying the next translation and/or rotation transformation. The 3-D rendered isosurfaces of the measured tomograms presented in Fig. 3a are indicated at the corresponding translation/rotation transformation. Scale bar indicates 5 μm.